\documentclass[%
 reprint,
 amsmath,amssymb,
 aps,
]{revtex4-2}

\usepackage{graphicx}
\usepackage{dcolumn}
\usepackage{bm}
\usepackage{multirow}
\usepackage{amssymb}
\usepackage{float}

\begin{document}

\title{Effect of $\alpha$-clusters on particle production in O$-$O and p$-$O collisions at LHC energies}
\author{Deependra Sharma}
\email{deep.phy@outlook.com}
\author{Arpit Singh}%
\email{arpit.ehep@gmail.com}
\author{Md. Samsul Islam}
\email{samsul.phy@gmail.com}
\author{Basanta Nandi}
\email{basanta@phy.iitb.ac.in}
\author{Sadhana Dash}
 \email{sadhana@phy.iitb.ac.in}
\affiliation{%
 Department of Physics, Indian Institute of Technology Bombay,\\
 Mumbai-400076, India 
}%

\begin{abstract}
In the present work, O$-$O collisions at $\sqrt{s_{NN}}$ = 7 TeV and p$-$O collisions at $\sqrt{s_{NN}}$ 
= 9.9 TeV are studied using PYTHIA8/Angantyr model for heavy-ion collisions. The theoretically predicted $\alpha$-cluster structure of oxygen nucleus is implemented in the model to investigate the effect of initial configuration of oxygen nucleus on final state observables. The results obtained from $\alpha$-cluster structure are compared with those obtained from Woods-Saxon nuclear charge density distribution. The Angantyr model simulation showed that the radial distribution of oxygen nucleus in $\alpha$-cluster configuration is more compact in comparison to the Woods-Saxon distribution. The results on charged and identified particle pseudorapidity distribution is obtained in the two initial state configuration of the oxygen nucleus. The results demonstrated that the effect of initial geometrical configuration is more distinct in the non-central collisions in comparison to the central collisions for both O$-$O and p$-$O collisions. 
\end{abstract}

\maketitle

\section{\label{sec:level1}Introduction}
Over the last two decades, the experiments at the Relativistic Heavy Ion Collider (RHIC) and the Large Hadron Collider (LHC) have provided a wealth of information on the properties of deconfined state of quarks and gluons, known as Quark Gluon Plasma (QGP), formed in ultra 
relativistic heavy-ion collisions~\cite{1,2,3,4,5}. These experiments have revealed that the QGP evolves hydrodynamically as a near perfect fluid with low viscosity to entropy density ratio ($\eta/s$)
~\cite{6,7,8}. These experiments have also revealed a heavy-ion like particle production patterns in 
small systems like p-p, and p-A collisions at the RHIC and the LHC energies~\cite{9,11,12,13,14,15,16}. Recently started LHC Run-3 will enhance our understanding about the observation of QGP like effects in small systems by providing data on various colliding systems with different center-of-mass energies. Most importantly, the data on oxygen-oxygen (O$-$O) and proton-oxygen (p$-$O) collisions at 7 TeV and 9.9 TeV center-of-mass energies, respectively, will shed light on the emergence of collective phenomenon and presence of partonic energy loss in small systems. Moreover, p$-$O collisions will provide constraints on nuclear parton distribution functions and small-x studies~\cite{17} along with aiding the cosmic ray modeling by providing valuable hadronic data which might provide solutions to the Muon Puzzle in ultra-high energy air showers and the ambiguity in the cosmic ray mass composition~\cite{18}. \par

Apart from comprehending the physics and bridging the gap between small and large system dynamics, 
the oxygen collisions will also provide a unique opportunity to test the existence of $\alpha$-cluster 
model of the oxygen nucleus. The existence of $\alpha$-cluster structure in elements having 4n nucleons, e.g. $^8$Be, $^{12}$C, $^{16}$O, etc., was first conceived by G. Gamow in 1930~\cite{19} due to the fact that $\alpha$ particles are highly stable with a large binding energy. The $\alpha$-cluster model for 4n-type nuclei has been further investigated by several authors~\cite{20} as an alternative to suitably describe the energy level structures which were not well explained by the shell model theories~\cite{21}. However, the configuration of $\alpha$-clusters in the oxygen nucleus and other light nuclei is still not well understood. Several model calculations support linear chain structure with four $\alpha$-clusters in the oxygen nucleus ~\cite{22}. A few model calculations suggest a tetrahedral structure of oxygen above the ground state~\cite{23,24}. However, few recent model calculations with chiral nuclear effective field theory~\cite{25}, skyrme energy density
functional theory~\cite{26}, and an algebraic cluster model~\cite{27} support the tetrahedral $\alpha$ configuration in the ground state of oxygen. Moreover, the consequences of square and kite shaped $\alpha$-cluster configurations in light nuclei has also been discussed in the literature~\cite{28,29}. \par

It has already been established through the collisions of deformed nuclei, such as $^{238}$U, and 
$^{129}$Xe, that the initial geometry of the colliding nuclei plays a crucial role in the final state 
effects~\cite{30,31,32,33,34,35,36}. The nuclear charge distribution affects the initial energy density deposited at the collision vertex, which in turn affects the particles production in the final state. Therefore, the presence of $\alpha$-clusters in the oxygen nucleus will manifest itself into the final state observables in the collisions of oxygen nucleus. Moreover, some recent calculations indicate towards imaging the shapes of atomic nuclei through collective flow assistance~\cite{37}, and selection of events based on $\langle p_{T} \rangle$ to isolate body-body collisions in deformed nuclei~\cite{38} will allow to understand the initial conditions and evolution of nuclear 
structure in high energy collisions.\par 

\begin{figure*}
    \centering
    \includegraphics[width=\textwidth]{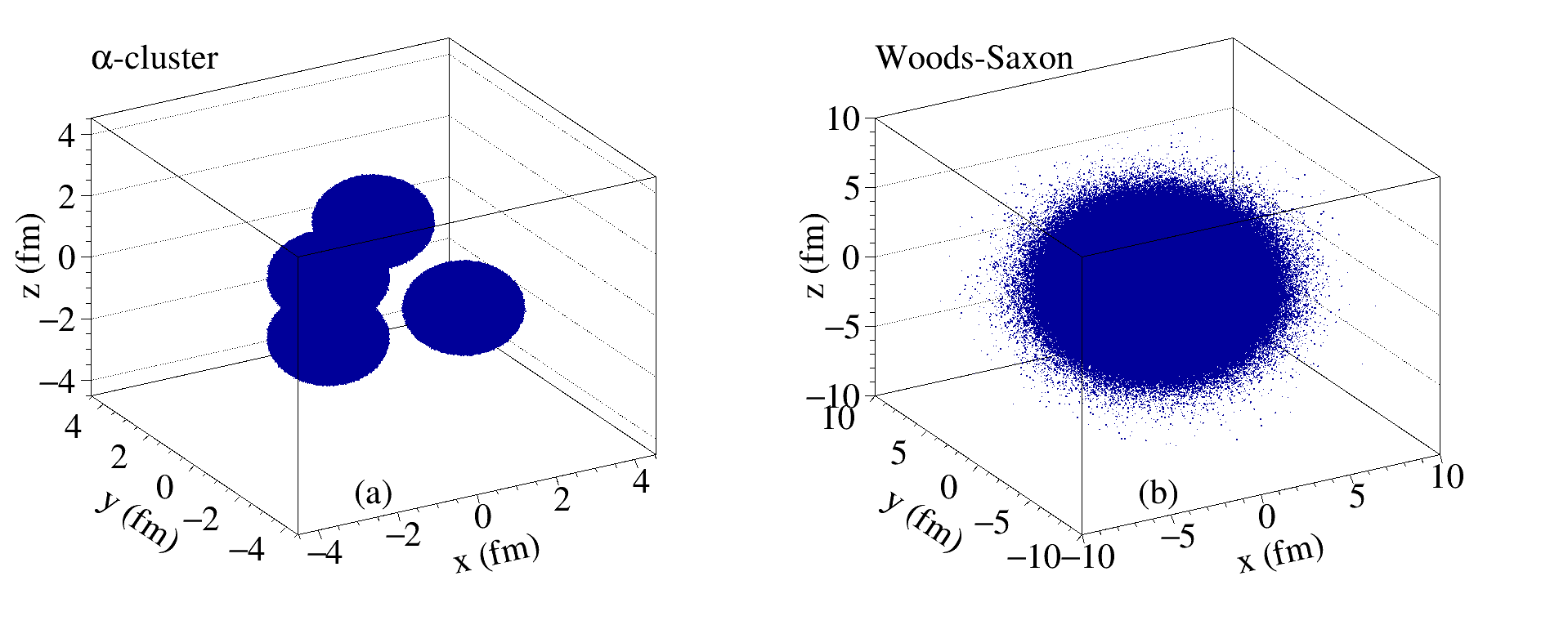}
    \caption{\label{fig:1a} A three-dimensional view of the (a)$\alpha$-cluster structure, and 
    (b) Woods-Saxon nuclear charge density distribution for $^{16}$O nucleus.}

\end{figure*} 

\begin{figure*}
    \centering
    \includegraphics[width=\textwidth]{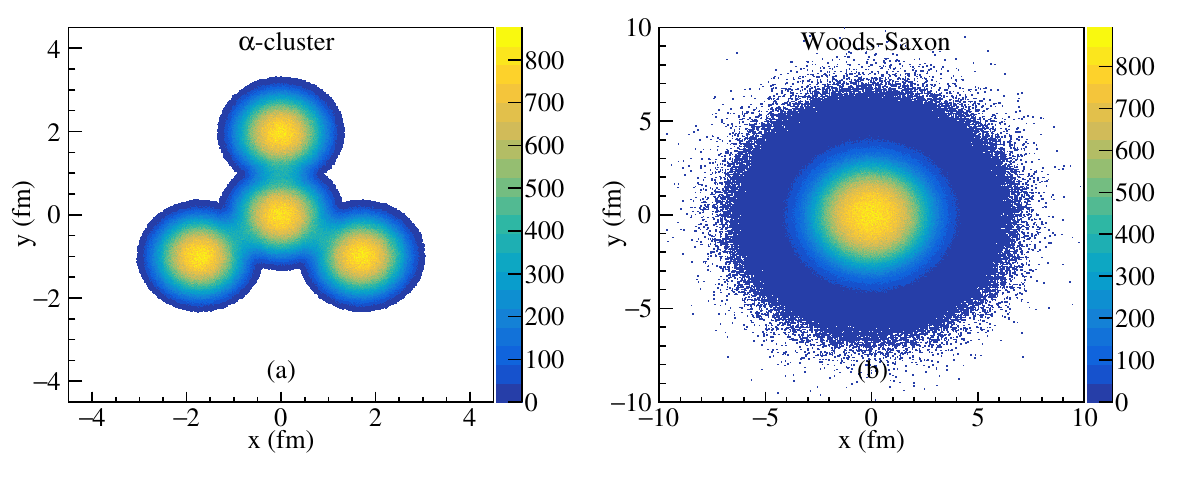}
    \caption{\label{fig:1b} Distribution of nucleons in the transverse (x-y) plane for (a) $\alpha$-cluster
    structure, and (b) Woods-Saxon nuclear charge density distribution for $^{16}$O nucleus.}
\end{figure*}

In this article, we investigate the effects of $\alpha$-clusters on particle production in O$-$O and p$-$O
collisions at $\sqrt{s_{NN}}$ = 7 TeV and 9.9 TeV, respectively. We geometrically incorporate 
$\alpha$-clusters in tetrahedral configuration in the oxygen nucleus in the PYTHIA/Angantyr event generator. To realize the consequences of $\alpha$-clusters on particle production, we compare the particle production in the O$-$O and p$-$O collisions without $\alpha$-clusters using standard Woods-Saxon nuclear charge density distribution. Since, the Angantyr model at present does not include hydrodynamic evolution, therefore, we focus on the final state observables such as particle multiplicity, and transverse momentum spectra of identified particles to discuss the effects of $\alpha$-cluster in O$-$O and p$-$O collisions at LHC energies. The rest of the article is organized as follows: Section II concisely describes the PYTHIA/Angantyr model and the implementation of $\alpha$-cluster structure in oxygen nucleus, Section III presents the results and discussions, and Section IV concludes the article. \par

\begin{figure}
    \centering
    \includegraphics[width=\linewidth]{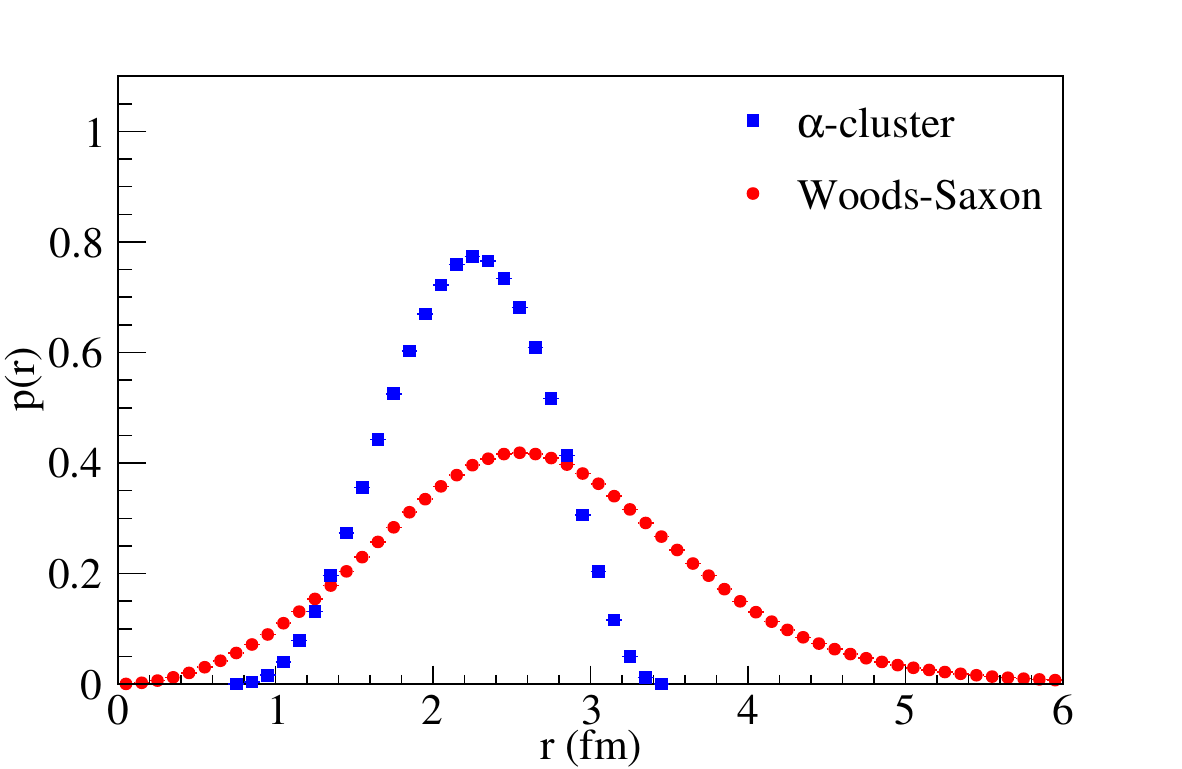}
    \caption{\label{fig:1c} Radial profile for distribution of the nucleons in the $^{16}$O nucleus for
    $\alpha$-cluster structure and Woods-Saxon nuclear charge density distribution.}
\end{figure}

\begin{table*}
    \caption{The impact parameter and average number of participating nucleons for O$-$O and p$-$O collisions at $\sqrt{s_{NN}}$ = 7 TeV 
    and 9.9 TeV, respectively, in different centrality intervals for $\alpha$-cluster and Woods-Saxon distributions.}
    \label{tab1}
    \resizebox{\textwidth}{!}{\begin{tabular}{|c|c|c|c|c|c|c|c|c|c|c|c|c|}
    
    \hline
     & \multicolumn{6}{c|}{O$-$O, $\sqrt{s_{NN}}$ = 7 TeV} & \multicolumn{6}{c|}{p$-$O, $\sqrt{s_{NN}}$ = 9.9 TeV} \\
    \hline
    \textbf{Centrality (\%)} & \multicolumn{3}{c|}{\textbf{Woods-Saxon}} & \multicolumn{3}{c|}{\textbf{$\alpha$-cluster}} & \multicolumn{3}{c|}{\textbf{Woods-Saxon}} & \multicolumn{3}{c|}{\textbf{$\alpha$-cluster}} \\
    \hline
    &$b_{min}$ (fm) &$b_{max}$ (fm) &$\langle N_{part}\rangle \pm$ rms &$b_{min}$ (fm) &$b_{max}$ (fm)& $\langle N_{part}\rangle \pm$ rms &$b_{min}$ (fm) &$b_{max}$ (fm)&$\langle N_{part}\rangle \pm$ rms &$b_{min}$ (fm) &$b_{max}$ (fm)& $\langle N_{part}\rangle \pm$ rms\\
    \hline
    0-5    &0 &1.40 &26.14 $\pm$ 2.53 &0 &1.04 &27.37 $\pm$ 2.34 &0 &0.84 &5.44 $\pm$ 2.64 &0 &0.74 &5.47 $\pm$ 2.66  \\ \hline
    5-10   &1.40 &2.00 &23.75 $\pm$ 2.71 &1.04 &1.48 &25.86 $\pm$ 2.50 &0.84 &1.19 &5.12 $\pm$ 2.48 &0.74 &1.06 &5.22 $\pm$ 2.52  \\ \hline
    10-20  &2.00 &2.84 &20.34 $\pm$ 3.11 &1.48 &2.13 &23.50 $\pm$ 2.79 &1.19 &1.70 &4.67 $\pm$ 2.26 &1.06 &1.51 &4.86 $\pm$ 2.35 \\ \hline
    20-40  &2.84 &4.11 &14.44 $\pm$ 3.80 &2.13 &3.16 &18.84 $\pm$ 3.45 &1.70 &2.46 &3.84 $\pm$ 1.93 &1.51 &2.21 &4.13 $\pm$ 2.05 \\ \hline
    40-60  &4.11 &5.17 &8.37 $\pm$ 3.45 &3.16 &4.10 &12.92 $\pm$ 3.70 &2.46 &3.14 &2.95 $\pm$ 1.55 &2.21 &2.82 &3.28 $\pm$ 1.70 \\ \hline
    60-80  &5.17 &6.29 &4.66 $\pm$ 2.58 &4.10 &5.14 &7.71 $\pm$ 3.47 &3.14 &3.92 &2.32 $\pm$ 1.24 &2.82 &3.53 &2.58 $\pm$ 1.38 \\ \hline
    80-100 &6.29 &19.95 &2.69 $\pm$ 1.54 &5.14 &11.00 &3.76 $\pm$ 2.35 &3.92 &13.41 &1.90 $\pm$ 0.95 &3.53 &10.76 &2.01 $\pm$ 1.04 \\ 
    \hline

\end{tabular}}
 
\end{table*}

\section{Model description}

\subsection{\label{sec:level2}Angantyr Model}
The Angantyr model is an extension of Pythia 8 event generator from p-p collisions to simulate nucleon-nucleus (p-A) and nucleus-nucleus (A-A) collisions without considering the formation of QGP matter~\cite{39}. The nucleon positions within the colliding nuclei are randomly sampled from the Woods-Saxon nuclear charge density distribution. The number of nucleon interactions and binary nucleon-nucleon (NN) collisions are obtained using Glauber formalism \cite{glauber1, glauber2}. The formalism is based on the Eikonal approximation in impact parameter space with the assumption that the projectile nucleons travel in straight lines and experience multiple sub-collisions with the target nucleons. The Glauber-Gribov colour fluctuation model is used to incorporate diffractive excitation corrections due to fluctuations in the nucleon substructure and extended to include cross-section fluctuations in target and projectile nucleon for p-A and A-A collisions. \par

The contribution from each nucleon-nucleon interactions to the final state is obtained using Fritiof model based on both non-diffractively and diffractively wounded nucleons which contribute to soft particle production. The events with multiple NN non-diffractive sub-collisions are further characterized into primary non-diffractive and secondary non-diffractive interactions based on impact parameter. Angantyr model uses full multi-parton interaction (MPI) machinery in Pythia 8 to generate non-diffractive, diffractive, and elastic p-p interactions. Later, all multiple sub-collisions are stacked together as colour singlet Lund strings which are then hadronized to produce a heavy-ion collision event.  \par

Pythia 8 model using the microscopic rope hadronization mechanism and QCD-based color reconnections (CR) for multi-parton interactions properly describes the production and enhancement of (multi-) strange hadrons in p-p collisions~\cite{47}. Angantyr model used in the current study uses MPI-based color reconnection without any implementation of rope hadronization mechanism. In the present scenario, for color reconnection mechanism the color dipoles that form the strings from different sub-collisions do not interact with each other. However, recent studies implemented the QCD-based color reconnection at hadronic level to understand naively the phenomenology of its effect on inter-sub-collision~\cite{48} and the rope hadronization mechanism to understand the enhanced strange meson and baryon production in heavy-ion collisions~\cite{49}.

\begin{figure*}
    \centering
    \includegraphics[width=\textwidth]{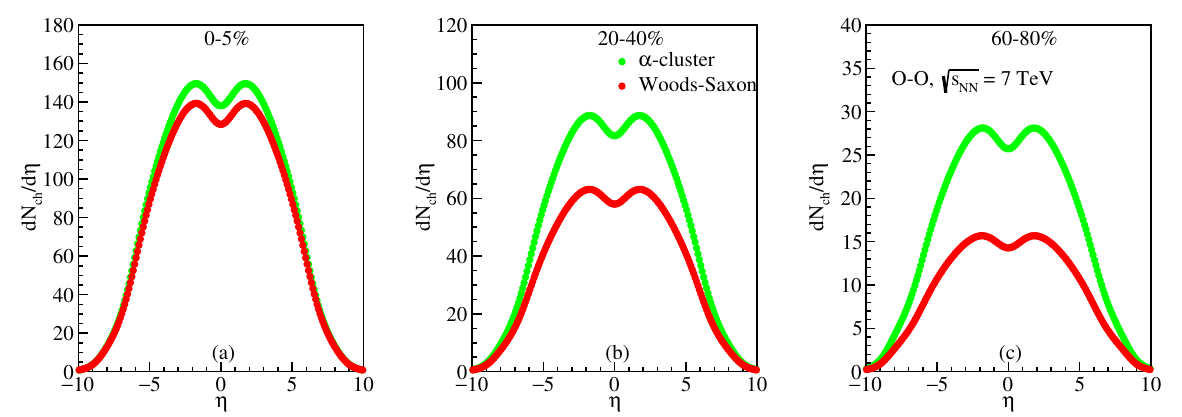}
    \caption{\label{fig:1} Pseudorapidity distribution of charged particles produced in 
    O$-$O collisions at $\sqrt{s_{NN}}$ = 7 TeV for (a) 0-5\%, (b) 20-40\%, and (c) 60-80\% 
    centrality intervals for $\alpha$-cluster structure and Woods-Saxon nuclear charge density
    distribution.}

\end{figure*}

\begin{figure*}
    \centering
    \includegraphics[width=\textwidth]{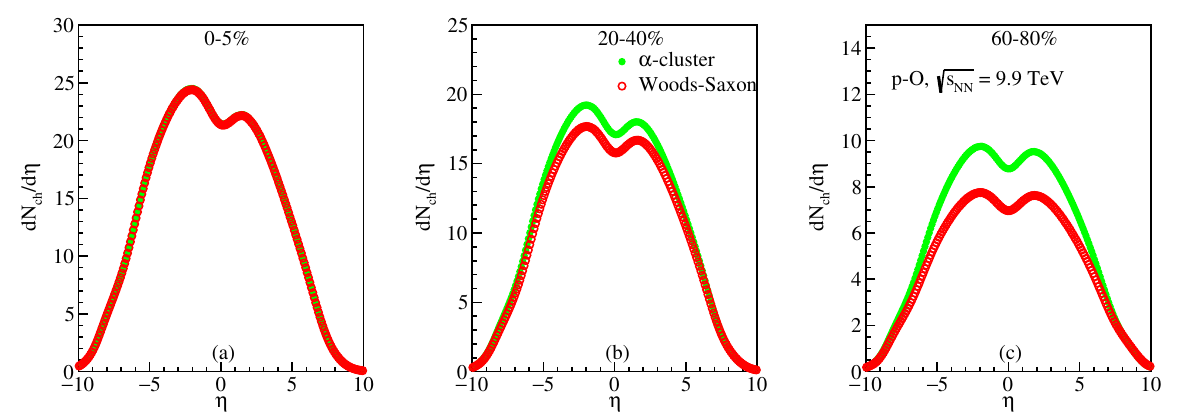}
    \caption{\label{fig:2} Pseudorapidity distribution of charged particles produced in 
    p$-$O collisions at $\sqrt{s_{NN}}$ = 9.9 TeV for (a) 0-5\%, (b) 20-40\%, and (c) 60-80\% 
    centrality intervals for $\alpha$-cluster structure and Woods-Saxon nuclear charge density
    distribution.}
\end{figure*}

\subsection{\label{sec:level3}$\alpha$-cluster structure implementation in $^{16}$O}
$^{16}$O nucleus consists of eight protons and eight neutrons which makes it doubly magic providing an exceptional nuclear stability. In light nuclei, such as $^{16}$O, the nucleons might rearrange themselves in a way to form $\alpha$-clusters and the nuclear mean field is not strong enough to break the clusters~\cite{20,40,41}. The tetrahedral structure with four $\alpha$-clusters in the oxygen nucleus has been provided by extended quantum molecular dynamics(EQMD) with the effective Pauli potential~\cite{42}. According to EQMD, the four $\alpha$-clusters are located at the vertices of a tetrahedron with the side length of 3.42 fm such that rms radius is 2.699 fm, which is in agreement with the experimental data of 2.6991 fm \cite{Oradius}. \par

The $\alpha$-cluster structure of $^{16}$O nucleus is incorporated in the Angantyr model geometrically by creating a regular tetrahedron with side length of 3.42 fm. For sampling the nucleon positions within the $\alpha$-clusters, the Woods-Saxon nuclear charge density is parameterized using a three-parameter Fermi (3pF) distribution. The 3pF distribution is given by, \par

\begin{equation}
    \rho(r) = \frac{\rho_{0}(1+w(\frac{r}{R})^{2})}{1 + \exp(\frac{r - R}{a})} ,
\end{equation} 
where $r$ is the radial distance of nucleon from the center of the nucleus, $\rho_{0}$ is the nuclear density at $r$ = 0, $R$ is the nuclear radius, $a$ is the diffuseness parameter, and $w$ is the skin thickness parameter. For each $\alpha$-cluster, $R$ = 0.964 fm, $a$ = 0.322 fm, and $w$ = 0.517 fm are used to obtain the nuclear charge density distribution~\cite{44}. The position of nucleons are randomized by applying three successive Euler rotations on event by event basis for the oxygen nucleus. To avoid overlapping, the minimum distance between two nucleons inside the same oxygen nucleus is kept larger than 0.9 fm which is the default hard core radius defined in Angantyr model. However, using this configuration in Angantyr model, a rms radius of 2.257$\pm$0.098 is obtained, which is a bit smaller than the predictions of EQMD.  

\begin{figure*}
    \centering
    \includegraphics[width=\textwidth]{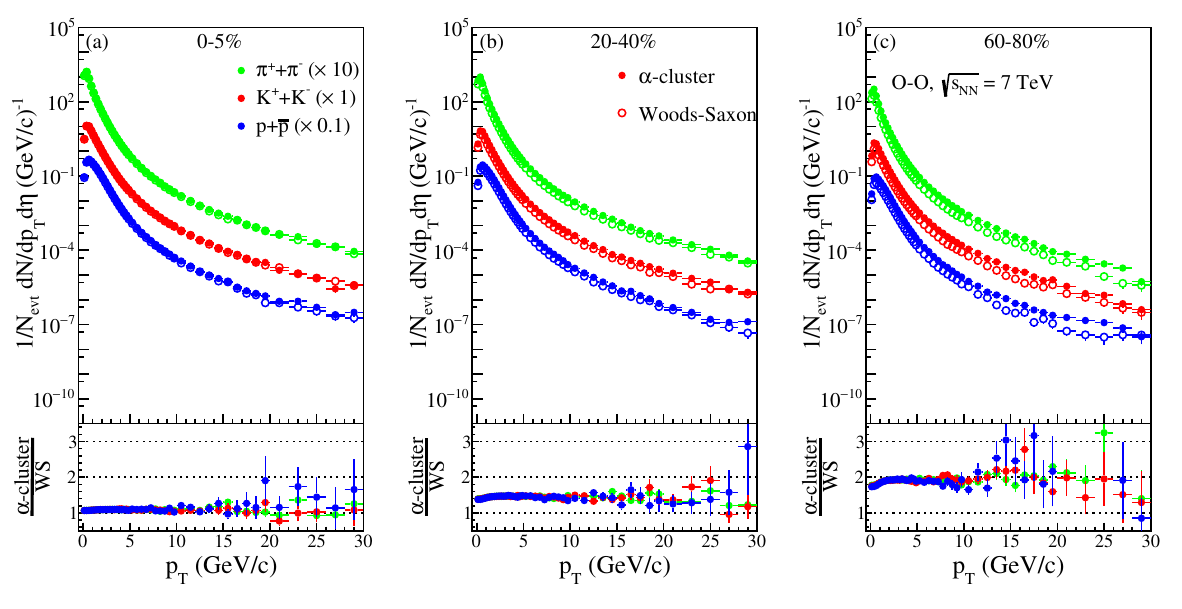}
    \caption{\label{fig:3} $p_T$-spectra of pions, kaons, and protons in O$-$O collisions at $\sqrt{s_{NN}}$ = 
    7 TeV in (a) 0-5\%, (b) 20-40\%, and (c) 60-80\% centrality intervals for $\alpha$-cluster and Woods-Saxon 
    distribution.}
\end{figure*}

\begin{figure*}
    \centering
    \includegraphics[width=\textwidth]{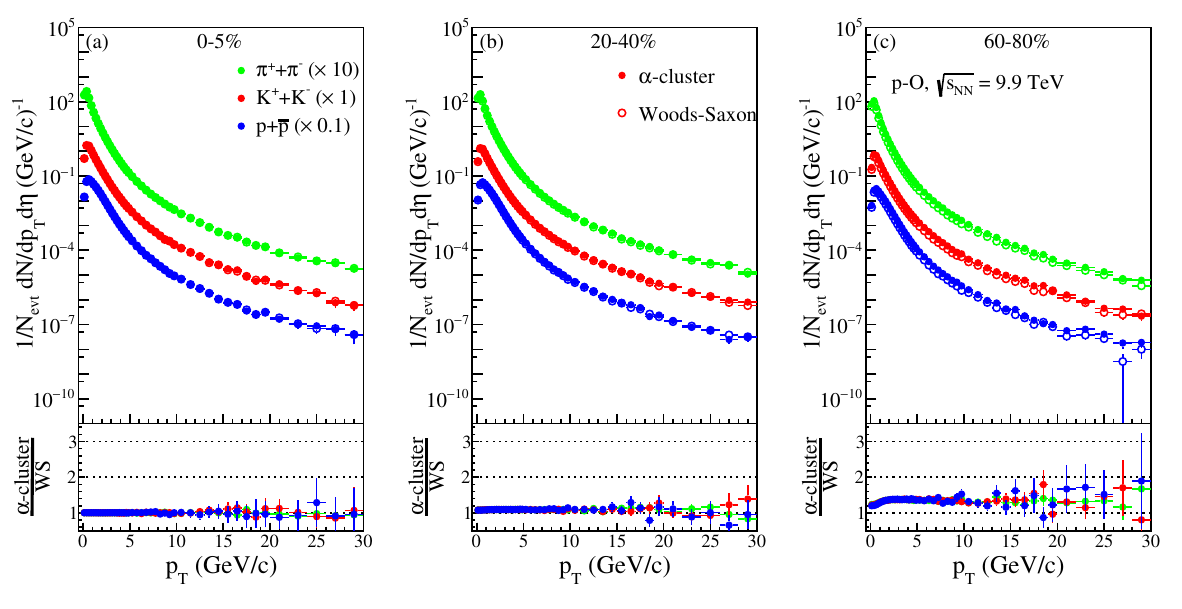}
    \caption{\label{fig:4} $p_T$-spectra of pions, kaons, and protons in p$-$O collisions at $\sqrt{s_{NN}}$ = 
    9.9 TeV in (a) 0-5\%, (b) 20-40\%, and (c) 60-80\% centrality intervals for $\alpha$-cluster and Woods-Saxon 
    distribution.}
\end{figure*}

\subsection{\label{sec:level4}Woods-Saxon Distribution}
In order to compare the effects of $\alpha$-cluster structure on particle production in O$-$O and p$-$O
collisions, the 3pF Woods-Saxon nuclear charge density distribution given in Eq. 1 is used for the oxygen nucleus. Here, $R$ = 2.608 fm, $a$ = 0.513 fm, and $w$ = $-$0.051 fm are used to obtain charge density profile for oxygen nucleus from which the nucleon positions are sampled~\cite{45}. Fig.~\ref{fig:1a} shows the three-dimensional view of the nucleon positions for $^{16}$O nucleus obtained from $\alpha$-cluster structure as well as the Woods-Saxon distribution. \par

The distribution  of nucleons in x-y plane for $\alpha$-cluster structure (left panel) and Woods-Saxon nuclear charge density distribution (right panel) for $^{16}$O nucleus is shown in Fig.~\ref{fig:1b}. It can be observed that the nucleons are more localized at the vertices of the tetrahedron in $\alpha$-cluster structure in comparison to the Woods-Saxon nuclear charge density distribution where nucleons are localized at the center of the oxygen nucleus. Fig.~\ref{fig:1c} shows the probability density of finding a nucleon inside the oxygen nucleus as a function of radial distance for $\alpha$-cluster structure and Woods-Saxon nuclear charge density distribution. The radial profile is more succinct for $\alpha$-cluster structure in comparison to Woods-Saxon nuclear charge density distribution which have a more smeared radial distribution. The radial profile shows that the probability density is maximum at the nuclear radius for both $\alpha$-cluster and Woods-Saxon distribution and the height of the probability distribution peak at the nuclear radius is about two times larger for $\alpha$-cluster structure in comparison to Woods-Saxon distribution.   \par

\section{\label{sec:level4}Results and Discussions}
In this section, the results obtained for the pseudorapidity ($\eta$) and transverse momentum ($p_{T}$) distributions
of identified particles produced in O$-$O and p$-$O collisions at $\sqrt{s_{NN}}$ = 7 TeV and 
$\sqrt{s_{NN}}$ = 9.9 TeV, respectively, are presented and discussed. The results for $\alpha$-cluster structure of oxygen nucleus are compared with the 3pF Woods-Saxon nuclear charge density distribution for 0-5\%, 20-40\%, and 60-80\% centrality intervals. The centrality intervals are determined using the impact parameter of the colliding nuclei. Table~\ref{tab1} shows the values of impact parameter
and average number of participating nucleons for O$-$O and p$-$O collisions at $\sqrt{s_{NN}}$ = 7 TeV  
and 9.9 TeV, respectively, in different centrality intervals for $\alpha$-cluster and Woods-Saxon distributions. 

\subsection{\label{sec:level:4.1} Pseudorapidity Distributions}
Fig.~\ref{fig:1} shows the $\eta$ distributions of charged particles produced in O$-$O
collisions at $\sqrt{s_{NN}}$ = 7 TeV for 0-5\%, 20-40\%, and 60-80\% centrality intervals. It can be clearly seen that the nuclear density has a significant effect on the $\eta$ distribution of charged particles produced in O$-$O collisions at $\sqrt{s_{NN}}$ = 7 TeV for all the centrality intervals. The charged particle 
pseudorapidity density for $\alpha$-cluster structure is higher than the Woods-Saxon nuclear
charge density distribution for all the centrality intervals. However, the difference becomes more pronounced as we move from central to peripheral collisions. This indicates that the $\alpha$-cluster oxygen nucleus deposits more energy in a smaller volume leading to higher energy density compared to the Woods-Saxon nuclear charge density distribution in the initial state. The higher initial energy density
manifests in larger production of particles in the final state. 
Similar studies using relativistic transport model~\cite{46} and relativistic hydrodynamics simulations~\cite{44} have observed an 
opposite trend in the pseudorapidity distributions of charged particles as a function of centrality.
Their results show a large difference in $\alpha$-cluster structure and Woods-Saxon nuclear charge 
density distribution for central collisions and relatively smaller difference for peripheral collisions. They also show that the $\alpha$-cluster structure and Woods-Saxon density profile 
has almost no dependence on charged particle pseudorapidity distribution at forward pseudorapidities.
However, the results obtained from the PYTHIA/Angantyr model show a significant difference in the
charged particle pseudorapidity distribution at forward pseudorapidities for $\alpha$-cluster
structure and Woods-Saxon nuclear charge density distribution for non-central collisions. \par

\begin{figure*}
    \centering
    \includegraphics[width=\textwidth]{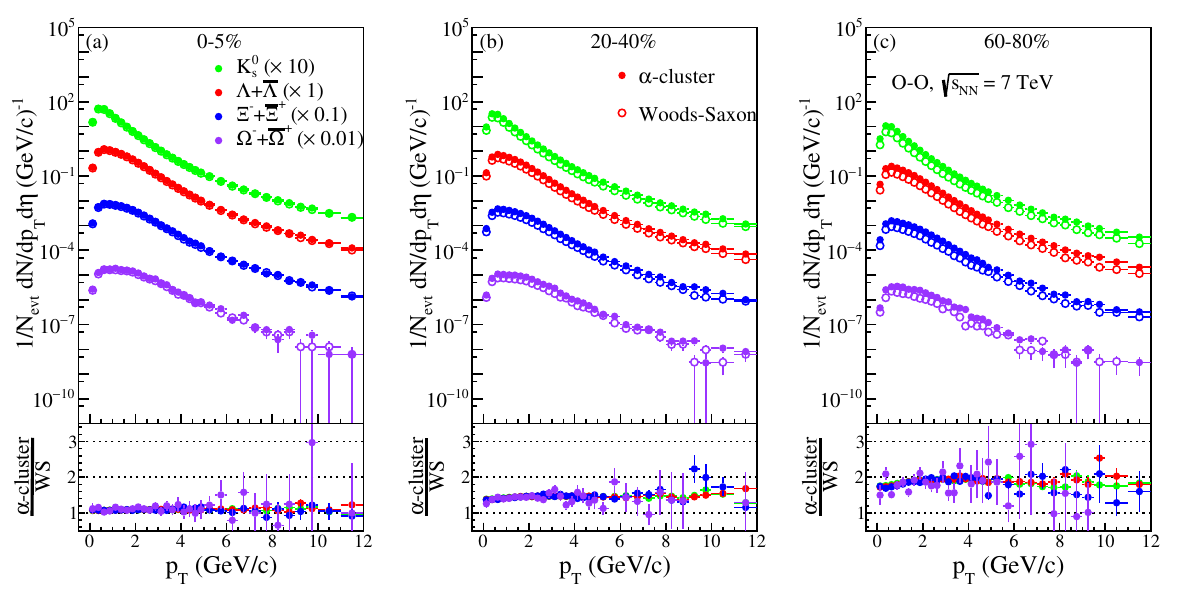}
    \caption{\label{fig:5} $p_T$-spectra of $K_{s}^{0}$, $\Lambda$, $\Xi$, and $\Omega$ in O$-$O collisions at $\sqrt{s_{NN}}$ = 
    7 TeV in (a) 0-5\%, (b) 20-40\%, and (c) 60-80\% centrality intervals for $\alpha$-cluster and Woods-Saxon 
    distribution.}
\end{figure*}

\begin{figure*}
    \centering
    \includegraphics[width=\textwidth]{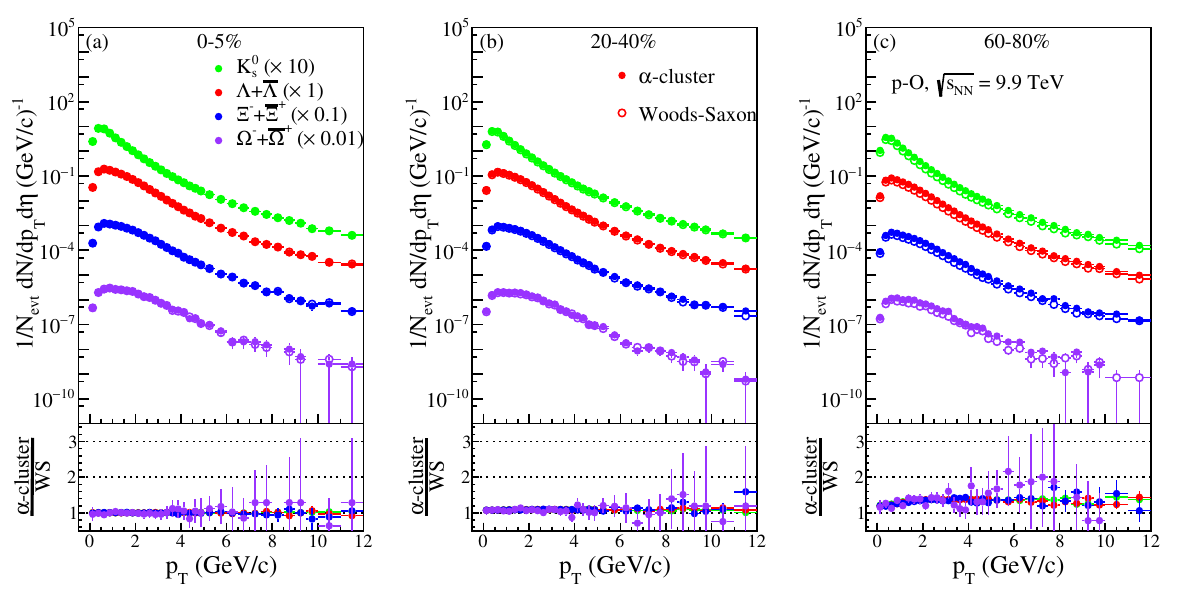}
    \caption{\label{fig:6} $p_T$-spectra of $K_{s}^{0}$, $\Lambda$, $\Xi$, and $\Omega$ in O$-$O collisions at $\sqrt{s_{NN}}$ = 
    9.9 TeV in (a) 0-5\%, (b) 20-40\%, and (c) 60-80\% centrality intervals for $\alpha$-cluster and Woods-Saxon 
    distribution.}    
\end{figure*}

Similarly, Fig.~\ref{fig:2} shows the pseudorapidity distributions of charged particles produced in 
p$-$O collisions at $\sqrt{s_{NN}}$ = 9.9 TeV for 0-5\%, 20-40\%, and 60-80\% centrality intervals. 
It is observed that the charged particle pseudorapidity distribution for 0-5\% centrality interval is almost 
independent of initial nuclear density profile.
However, there is significant difference for 20-40\% and 60-80\% centrality intervals between the  $\alpha$-cluster structure and the Woods-Saxon nuclear charge density distribution. It indicates that the energy density for both the 
configuration of oxygen nucleus is similar in central collisions. However, $\alpha$-cluster 
structure deposits more energy at the collision vertex than Woods-Saxon profile for non-central 
collisions, and hence, a relatively large particle production in non-central collisions for 
$\alpha$-cluster structure. A clear asymmetry in the charged particle pseudorapidity distribution 
is observed around $\eta$ = 0 for $\alpha$-cluster structure and Woods-Saxon nuclear charge 
density distribution for central and mid-central collisions. This is due to the fact that projectile 
proton probes a denser volume of the target oxygen nucleus increasing the particle yield in oxygen 
nucleus going in the beam direction. However,towards peripheral collisions, the charged particle 
pseudorapidity distribution for $\alpha$-cluster structure and Woods-Saxon nuclear charge density
becomes more symmetric around $\eta$ = 0. It may be due to the smaller number of participating 
nucleons and less dense interaction volume in the peripheral collisions, the situation becomes 
similar to the symmetric p-p collisions. \par 

The observation of higher charged particle pseudorapidity distribution for $\alpha$-cluster structure 
in non-central collisions in comparison to Woods-Saxon profile for both O$-$O and p$-$O collisions is 
quite obvious from $\langle N_{part} \rangle$ values shown in Table~\ref{tab1}. The $\langle N_{part} \rangle$ values for $\alpha$-cluster structure are consistently higher than the Woods-Saxon nuclear charge density distribution for all the centrality intervals in O$-$O and p$-$O collisions. As seen from Fig.~\ref{fig:1b}, the nucleons are more localized at the vertices of the tetrahedron in $\alpha$-cluster structure in comparison to the Woods-Saxon nuclear charge density distribution where nucleons are localized at the center of the oxygen nucleus. Therefore, even in non-central collisions, the collision system have larger nucleonic density in $\alpha$-cluster with respect to the Woods-Saxon profile.

\begin{figure*}
    \centering
    \includegraphics[width=\textwidth]{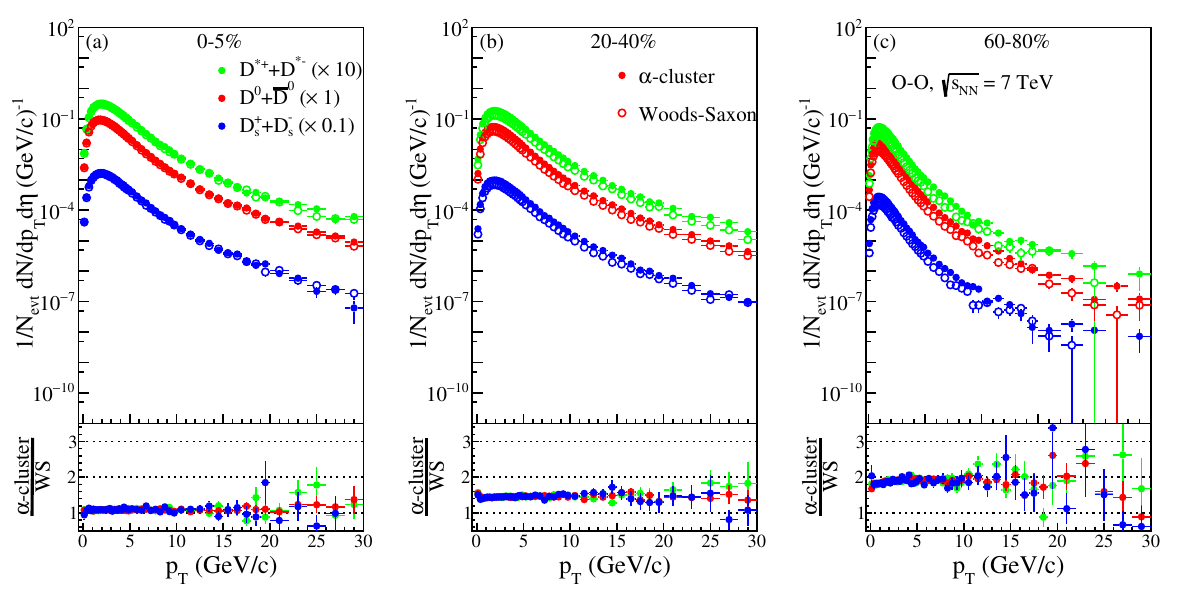}
    \caption{\label{fig:7} $p_T$-spectra of $D^{*}$, $D^{0}$, and $D_{s}$ in O$-$O collisions at $\sqrt{s_{NN}}$ = 
    7 TeV in (a) 0-5\%, (b) 20-40\%, and (c) 60-80\% centrality intervals for $\alpha$-cluster and Woods-Saxon 
    distribution.}
\end{figure*}

\begin{figure*}
    \centering
    \includegraphics[width=\textwidth]{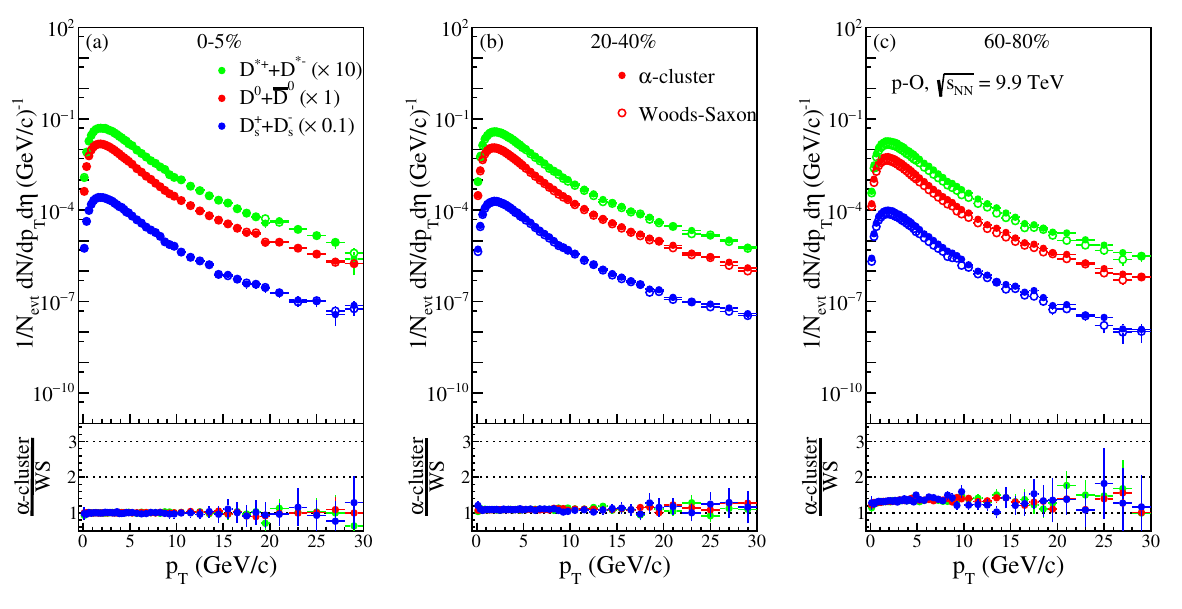}
    \caption{\label{fig:8} $p_T$-spectra of $D^{*}$, $D^{0}$, and $D_{s}$ in p$-$O collisions at $\sqrt{s_{NN}}$ = 
    9.9 TeV in (a) 0-5\%, (b) 20-40\%, and (c) 60-80\% centrality intervals for $\alpha$-cluster and Woods-Saxon 
    distribution.}
\end{figure*} 

\subsection{\label{sec:level:4.2} Transverse Momentum Distributions}
Fig.~\ref{fig:3} shows the transverse momentum distributions of pions ($\pi^{+}$ + $\pi^{-}$), kaons 
($K^{+}$ + $K^{-}$), and protons ($p$ + $\bar{p}$) produced in O$-$O collisions at $\sqrt{s_{NN}}$ = 
7 TeV for 0-5\%, 20-40\%, and 60-80\% centrality intervals in $|\eta|$ $<$ 0.8 for $\alpha$-cluster structure and Woods-Saxon charge density profile. The bottom panel of the figure shows the ratio of transverse momentum distributions of $\alpha$-cluster to Woods-Saxon profile. It can be observed that, in central collisions, the $p_{T}$-spectra of pions, kaons, and protons are similar for both the $\alpha$-cluster structure and Woods-Saxon nuclear charge density distribution as evident from the ratio which is 
consistent with unity. For non-central collisions, the $p_{T}$-spectra of pions, kaons, and protons hardens slightly for $\alpha$-cluster structure with respect to Woods-Saxon nuclear charge density
distribution. It indicates a relatively prominent overlap and subsequent hard scattering in $\alpha$-cluster 
structure in comparison to Woods-Saxon profile towards peripheral collisions.  \par

Fig.~\ref{fig:4} shows the $p_{T}$-spectra of pions ($\pi^{+}$ + $\pi^{-}$), kaons ($K^{+}$ + $K^{-}$), and protons ($p$ + $\bar{p}$) produced in p$-$O collisions at $\sqrt{s_{NN}}$ = 9.9 TeV for 0-5\%, 20-40\%, and 60-80\% centrality intervals in $|\eta|$ $<$ 0.8. The results for both $\alpha$-cluster structure and Woods-Saxon charge density profile are shown for comparison. The bottom panel of 
the figures show the ratio of $\alpha$-cluster to Woods-Saxon $p_T$-spectra for pions, kaons, 
and protons. It can be seen that the $p_T$-spectra for pions, kaons, and protons show no visible
difference for $\alpha$-cluster structure and Woods-Saxon nuclear charge density distribution for 
central and semi-central collisions. However, for peripheral collisions, a slight hardening of the $p_T$-spectra is observed for $\alpha$-cluster
structure with respect to Woods-Saxon nuclear charge density distribution for all the considered particles.  \par

 The transverse momentum distribution of strange hadrons from Angantyr model with MPI-based CR in O$-$O and p$-$O collisions at $\sqrt{s_{NN}}$ = 7 TeV and 9.9 TeV are obtained to understand the effect of initial density profile. Figs.~\ref{fig:5} and~\ref{fig:6} show the $p_{T}$-spectra of $K_{s}^{0}$, $\Lambda + \overline{\Lambda}$, $\Xi^{-} + \overline{\Xi}^{+}$, and $\Omega^{-} + \overline{\Omega}^{+}$ for O$-$O and p$-$O collisions, respectively, in different centrality intervals. The results obtained with $\alpha$-cluster structure are compared with the Woods-Saxon distribution. An observation similar to Figs.~\ref{fig:3} and~\ref{fig:4} is observed with higher particle yield and hardening of $p_{T}$-spectra 
for non-central collisions in $\alpha$-cluster structure of oxygen in comparison to Woods-Saxon 
distribution for both the collision systems. Since, the production of strange hadrons ( and light hadrons) in Pythia basically depends on the effective tunneling mechanism with string tension ($\kappa$) during hadronization through Lund fragmentation function, the absence of QCD-based CR, and rope hadronization mechanisms in Angantyr will not be able to provide sufficient string tension to enhance the production of strange mesons and baryons in heavy-ion collisions. \par

Unlike the light quarks, the heavy quarks in Pythia 8 are produced only in the hard scatterings and the partons showers~\cite{50}. The hard scattering processes like $qq \rightarrow Q\bar{Q}$ and $gg \rightarrow Q\bar{Q}$ are primary responsible for heavy quark production along with weak decay, Higgs decay, and top and bottom quark decay. The parton showers mainly produce heavy quarks through pair creation, flavour excitation or gluon splitting. The heavy quark masses are used for the perturbative description of their production which affects the matrix elements, splitting kernals, and the phase space of the heavy quark production coss-section. Figs.~\ref{fig:7} and~\ref{fig:8} show the $p_{T}$-spectra of $D^{*+} + D^{*-}$, $D^{0} + \overline{D}^{0}$, and $D_{s}^{+} + D_{s}^{-}$ obtained for $\alpha$-cluster structure and Woods-Saxon distribution in O$-$O and p$-$O collisions at $\sqrt{s_{NN}}$ = 7 TeV and 9.9 TeV, respectively, for different centrality intervals. Here also, an enhancement in D-meson yields and slight hardening of spectra in non-central collisions is observed in $\alpha$-cluster configuration relative to Woods-Saxon distribution for both the collision systems. 
 
\section{\label{sec:level5}Conclusions}
In this article, the effects of initial geometrical configuration of oxygen nucleus on the final state 
multi-particle production in O$-$O and p$-$O collisions at $\sqrt{s_{NN}}$ = 7 TeV and 9.9 TeV is explored
under the framework of Angantyr model for heavy-ion collisions. The nucleons in oxygen nucleus are expected to have a $\alpha$-cluster structure with four $^4$He ($\alpha$) nuclei sitting at each vertex of regular tetrahedron. The $\alpha$-cluster configuration of oxygen is implemented geometrically in the Angantyr model and final state observables like psuedorapidity distribution of charged particles and transverse momentum distribution of identified particles, such as final state charged particle, strange hadron and open heavy flavours are obtained. In order to realize 
the consequences of $\alpha$-clusters on final state observables, the commonly used 3pF Woods-Saxon charge density distribution in heavy-ion collisions is also utilized in the initial state of the Angantyr model. It is observed that the effect of $\alpha$-clusters is more pronounced in non-central collisions around mid and forward rapidites in comparison to Woods-Saxon distribution for both O$-$O and p$-$O collisions. It indicates that the study of particles at forward rapidities, such as $J/\psi$, $\Upsilon$, beam remnants, etc., may provide a better understanding of gluon saturation other QCD phenomena including the information about QGP and cold nuclear matter effects and most importantly the information about the structure of oxygen nucleus in the initial state. \par

The transverse momentum distribution of identified particles show an increase in particle yield and slight hardening of the particle spectra in non-central collisions for $\alpha$-clusters in comparison to Woods-Saxon distribution of the oxygen nucleus. The distinction between the two initial configurations on the final state observables may become more evident upon incorporating the QCD-based color reconnection mechanism, and rope hadronization for string fragmentation in the Angantyr model. The effect of these mechanisms on particle production in O$-$O and p$-$O collisions will be explored in our future work. 

\section{Acknowledgement}
This work has been supported by Anusandhan National Research Foundation (ANRF), Government of India, under grant No. RD/0122-SERBF30-001.  

\nocite{*}

\bibliography{main}

\end{document}